\documentclass[runningheads]{llncs}
\usepackage{graphicx}
\usepackage[frozencache=true,cachedir=.]{minted}
\begin{document}
\title{Diplomat: A conversational agent framework for goal-oriented group discussion\thanks{This work was supported by the United States Office of Naval Research under Contract N000141812767 and partially funded by an NSF REU grant, REU-CAAR, 
CNS-1952352.}}
\titlerunning{Diplomat: A conversational agent framework}
%
\author{Kevin Hogan \and
Annabelle Baer \and
James Purtilo}
\authorrunning{K. Hogan et al.}
%
\institute{Department of Computer Science, University of Maryland, College Park, MD 20742, USA}


\maketitle 
\begin{abstract}
Recent work in human-computer interaction has explored the use of conversational agents as facilitators for group goal-oriented discussions. Inspired by this work and by the apparent lack of tooling available to support it, we created Diplomat, a Python-based framework for building conversational agent facilitators. Diplomat is designed to support simple specification of agent functionality as well as customizable integration with online chat services. We document a preliminary user study we conducted to help inform the design of Diplomat. We also describe the architecture, capabilities, and limitations of our tool, which we have shared on GitHub\footnote{https://github.com/kevin-hogan/diplomat}.

\end{abstract}

\section{Introduction}
\label{sec:Introduction}

Group discussion is a cornerstone of human collaboration. Unlike a lecture or dictation, it is a medium of information exchange which allows \textit{all} group members to freely and candidly express their ideas and opinions. Unfortunately, the same characteristic of group discussion that promotes this ideal, namely its organic and unstructured nature, renders it entirely vulnerable to misaligned behavior. We suspect most readers have been a party to a discussion gone haywire, perhaps due to an emotional disagreement or a particularly charismatic discussion participant.

It comes as no surprise, therefore, that the annals of scientific research have, for the past century, been littered with contributions to the understanding and improvement of group discussion. Echoing Thorndike's assessment \cite{thorndike1938effect}, we identify M{\"u}nsterberg as perhaps \textit{the} pioneer in this research area with his early work on jury deliberations \cite{munsterberg1914mind}. Selected works between then and now have explored the comparison of individual and group performance on problem-solving \cite{shaw1932comparison,lorge1958survey,laughlin2011group}, formal modeling of group decision-making \cite{davis1973group,brodbeck2007group}, and group creativity \cite{ziller1962group,mcleod1996ethnic,lamm1973group}.

As computer scientists, we're primarily interested in the application of computers towards the improvement of group discussion, a more modern research area that blossomed alongside personal computing in the late 1980s. Nunamaker et. al. were the first to introduce the term ``Electronic Meeting System" (perhaps a descendant or variant of the ``Group Communication Support System" \cite{pinsonneault1989impact}), which categorizes technologies ``designed to directly impact and change the behavior of groups" \cite{nunamaker1991electronic}. Over the past few decades, researchers have produced a plethora of systems \cite{van1997supporting,chandrasegaran2019talktraces,briggs1997meetings,siemon2019one} that fit this categorization (though usage of the term itself seems to have waned over the years).

Our work fits in this broad category as well, but can more specifically be classified as research seeking to automate the role of the facilitator in group goal-oriented discussion. We've long known of the benefits that facilitators add to group conversations \cite{viller1991group}, but the advent of advanced natural language interfaces to computation has spurred the development of conversational agents (CAs) to assist or replace these facilitators \cite{lee2020solutionchat,kim2020bot,shamekhi2018face,peng2019gremobot}. Beyond the obvious benefit of scale and reproducibility, a CA can also provide the unique benefit of being consistently unbiased in its facilitation (assuming it is programmed without bias, of course).

In our investigation of this area, we noticed an absence of software frameworks to develop CAs for group goal-oriented conversation. This stands in contrast to variety of frameworks available to develop CAs for dyadic conversations (i.e., virtual assistants and chatbots) \cite{daniel2020xatkit,janarthanam2017hands}. In this work, we seek to fill this gap, designing for Internet-based text communication due to the popularity of this medium and the frequency of CAs built around it. We make the following contributions:

\begin{enumerate}
    \item A preliminary user study exploring participant perceptions of conversational agents and reactions to specific features.
    \item Diplomat, an open-source Python-based framework for the development and deployment of CA discussion facilitators.
\end{enumerate}

The rest of the paper proceeds as follows: Section \ref{sec:RelatedWork} describes prior research related to each of the contributions listed above. Section \ref{sec:Experiment} describes the preliminary user study mentioned in our first listed contribution. Section \ref{sec:Diplomat} describes the architecture, implementation, and limitations of Diplomat. Section \ref{sec:FutureWork} discusses possible future research directions inspired by this work, and we conclude in Section \ref{sec:Conclusion}.
\section{Related Work}
\label{sec:RelatedWork}

We highlight related work in two areas (corresponding to each of our main contributions listed in the Introduction): that which examines the effect of a conversational agent in discussion and that which simplifies the development of conversational agents. Within the former area, we see an abundance of work in the context of dyadic, or one-on-one, conversation. XiaoIce, one of the world's most popular chatbots, has been studied for its ability to emotionally connect with users \cite{zhou2020design}. Ciechanowski et al. study the differences between user reactions to text-based and avatar-based versions of an agent \cite{ciechanowski2017necessity}. Zhao et al. study the mechanism for rapport building in dyadic interactions with a CA \cite{zhao2014towards}. We also see efforts in the context of group discussion, which pertains to our work more directly. CAs have been studied for their ability to aid facilitators \cite{lee2020solutionchat}, replace facilitators \cite{kim2020bot}, increase critical thinking \cite{goda2014conversation}, prime conversations \cite{isbister2000helper}, create arguments \cite{hamilton2018argument}, and improve conversation flow \cite{deloach2007effectively}. Our preliminary user study described in Section \ref{sec:Experiment} helps to validate a general conclusion found throughout this work: that conversational agents can improve group discussions and provide value to their participants.

Work aiming to simplify the development of conversational agents is heavily focused on the dyadic discussion context. Several solutions exist for the development of commercial chatbots, such as Microsoft BotFramework, IBM Watson, and Google DialogFlow. Xatkit is a more research-oriented framework based in model-driven engineering that provides a domain-specific language for the specification of chatbot rules and allows for flexible integration with intent-recognition providers \cite{daniel2020xatkit}. However, none of these frameworks are tailored towards group, goal-oriented discussion. The Neem Platform \cite{barthelmess2005neem} is a research test-bed for collaborative applications which includes support for ``virtual participants". However, this support appears to be a small part of a broader platform and there is no publicly-available implementation. Diplomat, in contrast, is entirely focused on supporting the development and deployment of conversational agents for group discussion. We've released the software as a tool for those who may benefit from this support in their research or other efforts.

\section{Preliminary User Study}
\label{sec:Experiment}

\subsection{Summary}

To explore the impact of a conversational agent in a goal-oriented group conversation, we set up an experiment in which five groups were instructed to chat about an opinion-oriented prompt and reach a conclusion. We used the Wizard-of-Oz method \cite{green1985rapid} so that a human researcher could participate in the discussion while masquerading as an automated agent. This method allowed us to examine the effect of an agent without the cost of its implementation. The researcher interacted in discussions according to the pre-defined ruleset described in Section \ref{sec:Experiment-Procedure}. In Section \ref{sec:Experiment-Conclusions} we discuss the takeaways from this study and how it motivated the design of Diplomat. This study was performed after being approved by the Institutional Review Board, University  of  Maryland, College Park with Ref. no. 1633038-1.

\subsection{Procedure}
\label{sec:Experiment-Procedure}
The experiment was held remotely and coordinated using video conferencing. We did not observe any uncontrollable variables such as internet connection or other distractions that played a role in the outcome of the experiment. Participants were undergraduate volunteers randomly (with the exception of some scheduling constraints) split into five groups of four or five students each. Following from our selection of Internet-based chat as our preferred communication medium (see Section \ref{sec:Introduction}), we chose to host participant discussions on the popular chat provider, Slack. Each group discussed in a separate channel, and was unable to see the channels for other groups. At the beginning of the session, the moderator (under the identity of an autonomous agent) announced a question prompt and instructed participants to discuss this topic in their channel for 20 minutes (or until reaching consensus, if this happened sooner). At the end of the session, the participants filled out a survey via Google Forms answering questions about their feelings towards the conversation.

Our ``agent" interacted with participants differently for each conversation. The rules for each of the five scenarios are presented in the list below. We designed the rules based on our intuitions about what would promote constructive and fair deliberations.

\begin{itemize}
    \item \textbf{Information}: The agent will send new links and suggest new topics to the users when there is a lull (2 minute silence) in conversation. These links were predetermined by taking the first 5 links from a Google search of the full text of the prompt.
    \item \textbf{Timing}: The conversation will be actively timed (20 minutes). The agent will give time warning signals when there are 10, 5, and 2 minutes remaining.
    \item \textbf{Under/Overspeaking Notification}: The agent will address users who haven’t spoken in 8 messages after speaking their last message. The agent will address users who have spoken over ½ the messages in a grouping of 8 or more messages.
    \item \textbf{Rules 1-3 combined}: All rules apply
    \item \textbf{Control}: The agent will not be a part of the control conversation
\end{itemize}

Using a post-experiment survey, we measured the users’ 1) satisfaction with the outcome of the meeting and 2) reactions to the agent. Kim et al. reports inconsistent user attitudes as a result of the study, but states user testimonies that support the idea that the conversational agent removed “relational burdens” by addressing users’ participation as an objective measure \cite{kim2020bot}. We wanted to understand if our intervention scenarios could similarly reduce burdens. We also measured the frequency of expected behaviors in response to agent interventions (we omit the Timing intervention because, after the experiment, we realized the difficulty in articulating and measuring an expected response from this intervention). These expected behaviors are listed below:

\begin{itemize}
    \item \textbf{Information}
    \begin{itemize}
        \item People respond to link with insights clearly taken from the suggested link.
    \end{itemize}
     \item \textbf{Underspeaking Notification}
    \begin{itemize}
        \item No direct response, but the person says a new message within the next 5 messages. 5 messages was chosen because on average, each conversation had 5 people, giving the user a small window of time to think about and message their response. 
    \end{itemize}
    \item \textbf{Overspeaking Notification}
    \begin{itemize}
        \item No direct response, but the person stops responding for any period of time greater than 5 messages for reasons stated above.
    \end{itemize}
\end{itemize}

\subsection{Results}

As mentioned in the previous subsection, the post-experiment survey asked for both satisfaction with the outcome of the meeting and the reactions of participants to the agent. With one exception across all participants, participants expressed satisfaction with the meeting outcome. Therefore, we don't observe any substantial differences in satisfaction across the various rule scenarios we tested. The participants had expressed mixed reactions to the agent. Roughly 50\% of all participants (excluding the control group) indicated in a free-form response that they would choose to include the agent in another conversation. However, about 25\% expressed a negative impression of the agent, calling it ``annoying" and ``not super useful". The remainder were neutral. Participants assigned an agent in the Under/Overspeaking scenario responded most positively, with 2 positive and 2 neutral reactions. Those assigned to an agent in the Information scenario were most critical of the agent.

For each of the expected behaviors, the majority of users responded as expected. There two cases across all experiments where the agent addressed a user for overspeaking. In one case, the user only sent a total of two more messages in the remaining third of the conversation. The other user initially sent less frequent messages, but eventually returned to a higher frequency by the end of the conversation. There were three cases of underspeaking. In all of the cases, the notification happened so close to the end of the conversation, we do not have sufficient data to observe how the agent impacted the users' participation in the long run. However, in each of these instances, the user sent at least one message after the notification from the agent.

\subsection{Takeaways}
\label{sec:Experiment-Conclusions}

We caution against drawing strong conclusions from the experiment presented for several reasons. First, this was a a small study with only enough participants to conduct one group discussion for each of the tested scenarios. Without repeated trials in each scenario, it is impossible to discern whether variations in outcomes and perceptions across groups are caused by agent intervention rules or instead by the differences in participant opinions and tendencies. Second, intervention rule activation is infrequent by design, which, while sensible in preventing the agent from overwhelming the conversation, makes it difficult to analyze the typical effect of a given intervention type. Finally, participants were prompted to discuss a topic on which they may have had varying levels of interest and prior knowledge. We believe that additional framing and context of the conversation topic may lead to more consistent engagement across participants.

We share this experiment in spite of these concerns because we maintain that there are lessons to be gleaned from it. The results motivate a larger-scale experiment to alleviate some of the aforementioned concerns. This is precisely the driving force behind our decision to build Diplomat (described in the next section), which we see as an enabler of scale (the Wizard-of-Oz method, which requires the participation of a trained operator, is not as scalable). Also, our results indicate that even very crude intervention rules such as those we tested in this study are viewed as valuable by some participants. This motivates further research into more advanced or tunable interventions that may provide much more value to a group discussion.

\section{Diplomat}
\label{sec:Diplomat}

\subsection{System Design}

As described in Section \ref{sec:Experiment-Conclusions}, we were inspired to develop a framework that would facilitate the specification, configuration, and composition of the sort of conversational agent interventions we tested. Like the human ``agent" from our Wizard-of-Oz experiment, agents implemented in our tool communicate with groups via Internet-based chat discussions.

We based the design of our tool, which we've called Diplomat, on the following definitions and principles.
\begin{itemize}
    \item Definitions
    \begin{itemize}
        \item A \textit{transcript} represents the entire history of a group chat conversation consisting of all messages, the authorship of those messages, and the timestamps at which those messages were received.
        \item An \textit{intervention} is any interaction in the chat interface generated by the agent (currently, a message is the only supported intervention type).
        \item A \textit{reactive intervention} is an intervention triggered by user activity (e.g., our ``Overspeaking" intervention)
        \item A \textit{passive intervention} is an intervention triggered only by the passage of time (e.g., our ``Timing" intervention)
        \item An \textit{agent feature} is a deterministic rule that can be expressed as a function from a transcript to a list of interventions.
        \item A \textit{feature configuration} is a set of parameters that can be modified to tweak the behavior of a chatbot feature (e.g., the ``window size" for our overspeaking feature).
    \end{itemize}
    \item Principles
    \begin{enumerate}
        \item Agent features can be combined in a single agent, but should operate independently of one another.
        \item The framework should not be coupled with any particular chat service. It should, however, provide a straightforward method for integration with these services.
        \item The programming model for features should be simple and modular
    \end{enumerate}
\end{itemize}

Our solution is a Python-based framework that allows for the specification of agent features in the form of a subclass. In Figure \ref{fig:base_class} we see the base class for agent features. A feature subclass must implement a constructor which receives a feature configuration dictionary (loaded by the framework from a \texttt{JSON} file) and must produce an instance of the subclass parameterized by the configuration. It also must implement \texttt{generate\_interventions}, which (consistent with the agent feature definition above) receives a transcript and returns a list of interventions. Diplomat guides its users towards the fulfillment of Principle 1 by requiring the implementation of each feature in a separate subclass. This requirement also encourages the modularity referred to by Principle 3. As Principle 1 mentions, a particular instance of an agent may consist of several features. Whether or not a feature is enabled for an agent is determined by the presence or absence of a configuration for that feature in the \texttt{JSON} file that specifies the full desired configuration for an agent. As an additional note regarding feature specification, we highlight the fact that the \textit{entire} chat transcript is provided to \texttt{generate\_interventions}. This allows for stateless feature specification even for functions depending on long-term correspondence of chat data.

\begin{figure}[h]
\begin{minted}{python}

class FeaturePlugin(metaclass=abc.ABCMeta):
    @abc.abstractmethod
    def __init__(self, config: dict):
        pass

    @abc.abstractmethod
    def generate_interventions(
        self,
        chat_transcript: List[Message],
        author_id_for_chatbot) -> List[Message]:
        pass

\end{minted}
\caption{The abstract base class that users of Diplomat must extend to implement new conversational agent features.}
\label{fig:base_class}
\end{figure}

In fulfillment of Principle 2, the framework provides another base class that can be flexibly extended for integration with a variety of chat services (see Figure \ref{fig:base_class_integrate}). Users of the framework need not directly manage interactions with the rest of the framework when specifying the integration. Instead, they must only define the connection to the API from which a transcript is loaded, the conversion of that transcript to the internal representation, the conversion of interventions from the internal representation to the chat service API representation, and the post of interventions via the chat service API for writing messages. The constructor allows the user to tailor the polling interval to their liking.

\begin{figure}[h]
\begin{minted}{python}

class ChatServiceToBotIntegrator(metaclass=abc.ABCMeta):
    def __init__(self, path_to_config: str, seconds_per_poll: float):
        ...

    @abc.abstractmethod
    def request_transcript_and_convert_to_message_list(self) -> List[Message]:
        pass

    @abc.abstractmethod
    def post_agent_interventions(self, interventions: List[Message]) -> None:
        pass

\end{minted}
\caption{The abstract base class that users of Diplomat must extend to integrate with an external chat service.}
\label{fig:base_class_integrate}
\end{figure}

\begin{figure}[h]
\centering
\includegraphics[width=1\textwidth]{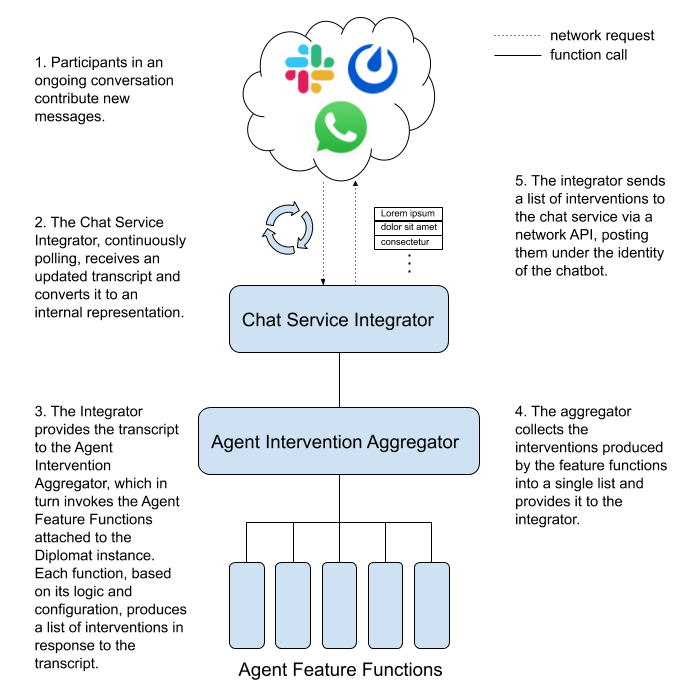}
\caption{Diplomat architecture diagram and intervention lifecycle}
\label{fig:lifecycle_diagram}
\end{figure}

Figure \ref{fig:lifecycle_diagram} illustrates the architecture of Diplomat and describes the lifecycle of an intervention in relation to the components of that architecture. We won't restate the phases of the lifecycle captured in the figure, but we will add some commentary and clarification here. The first two steps in the lifecycle surround the integration with external chat services described in the previous paragraph. The logos in the figure (for readers who may be unfamiliar) represent popular chat services. Diplomat provides an example integration to Slack, but does not provide additional integrations out-of-the-box. Another point of clarification is that Diplomat only support interventions in the form of messages from the agent. We leave the terminology general, however, as we hope to support additional forms of intervention in the future (emoji reaction, message replies, private messages, etc.) without modification to the overall architecture.

\subsection{Limitations}
\label{sec:Diplomat-Limitations}

Diplomat was designed with simplicity as a first-principle. The consequence is that the framework, especially in its nascent form, is limited in its performance, its expressiveness, and its accessibility. While the computational performance of Diplomat has not been analyzed thoroughly, some issues are apparent from its architecture. For example, Diplomat receives full conversation transcripts from chat services via polling. This means that Diplomat will redundantly load old conversation data continuously rather than keep track of a transcript locally (updating as new messages come in). To avoid potential complexities associated with state management, we opted for the simpler, polling approach for our initial release of the software. Beyond this, we note that only one Diplomat agent can be run per Python process, and that IO with the external chat service is synchronous. While introducing concurrency to the framework would allow for more efficient utilization of resources, it also introduces complexity that we chose to avoid undertaking for now. Due to these limitations, we recommend Diplomat for use in short conversations with modest numbers of participants.

Diplomat's expressiveness is limited due to the independence of agent features and the uniformity of interventions. A user may reasonably desire to develop a conversational agent in which features are dependent. The simplest example of this would be an agent that, at a given time, only produces one intervention corresponding to the feature that was most directly triggered by chat activity. Unfortunately, a Diplomat user would have to combine all dependent conceptual features into a single implemented feature to create an agent with this behavior. The uniformity of interventions (i.e., the fact that the only currently supported Diplomat intervention type is a message) has been described in the previous subsection.

When we mention the ``accessibility" of Diplomat, we mean its accessibility to those who wish the develop agents on top of the framework. Rather than provide an interface for agent specification that is abstracted from code, say via a domain-specific language or a drag-and-drop interface, our interface is the implementation of a Python class. This requires that users have programming skills. Also, while we provide some examples of chat service integration and agent features along with our software release, these examples are, as of this writing, not numerous. Users may have to invest significant time building agents with Diplomat.

Despite these limitations, we've released Diplomat with the optimism that it will provide value to those looking to develop conversational agents for group discussion and that it will, over time, develop into a more mature and robust framework.
\section{Future Work}
\label{sec:FutureWork}

We propose future research in two main thrusts: 1) conversational agent research at scale, and 2) improvement of Diplomat. In the first thrust, we're interested in multiple forms of ``scale". The most obvious form is scaling the participation in experiments. As described in Section \ref{sec:Experiment-Conclusions}, an abundance of data is essential to understanding the nuanced effect that a conversational agent can have on a group discussion. We also consider scaling the diversity of agent interactions. We plan to experiment with a much wider variety of intervention rules than those described in Section \ref{sec:Experiment-Procedure}. This may include simple rules with diverse configurations (e.g., tweaked thresholds for under/overspeaking, agents with permutations of a feature set) or more complex rules rooted in advanced NLP techniques (e.g., topic modeling, sentiment analysis).

Research into improving Diplomat will be based on addressing the limitations described in \ref{sec:Diplomat-Limitations}. We hope to cultivate a community of researchers interested in this tool and to continue maintaining it as an open-source project.
\section{Conclusion}
\label{sec:Conclusion}

In this paper, we introduced Diplomat, a Python-based framework for implementing conversational agents to be deployed in group, goal-oriented discussion. We also document a user study which, while preliminary and narrow, provides us with insights into how group discussion participants react to various interventions by a conversational agent. By investing in a framework like Diplomat, we hope to accelerate the development of conversational agents that can mitigate the familiar difficulties group discussion can too often present. The unprecedented collaborative capabilities with which the information age provides us will only be blessings if we have mechanisms to ensure their constructive and responsible use.

%
%
%
%

\bibliography{references.bib}{}

\begin{thebibliography}{10}
\providecommand{\url}[1]{\texttt{#1}}
\providecommand{\urlprefix}{URL }
\providecommand{\doi}[1]{https://doi.org/#1}

\bibitem{barthelmess2005neem}
Barthelmess, P., Ellis, C.A.: The neem platform: An evolvable framework for
  perceptual collaborative applications. Journal of Intelligent Information
  Systems  \textbf{25}(2),  207--240 (2005)

\bibitem{briggs1997meetings}
Briggs, R.O., De~Vreede, G.J.: Meetings of the future: enhancing group
  collaboration with group support systems. Creativity and Innovation
  Management  \textbf{6}(2),  106--116 (1997)

\bibitem{brodbeck2007group}
Brodbeck, F.C., Kerschreiter, R., Mojzisch, A., Schulz-Hardt, S.: Group
  decision making under conditions of distributed knowledge: The information
  asymmetries model. Academy of Management Review  \textbf{32}(2),  459--479
  (2007)

\bibitem{chandrasegaran2019talktraces}
Chandrasegaran, S., Bryan, C., Shidara, H., Chuang, T.Y., Ma, K.L.: Talktraces:
  real-time capture and visualization of verbal content in meetings. In:
  Proceedings of the 2019 CHI Conference on Human Factors in Computing Systems.
  pp. 1--14 (2019)

\bibitem{ciechanowski2017necessity}
Ciechanowski, L., Przegalinska, A., Wegner, K.: The necessity of new paradigms
  in measuring human-chatbot interaction. In: International Conference on
  Applied Human Factors and Ergonomics. pp. 205--214. Springer (2017)

\bibitem{daniel2020xatkit}
Daniel, G., Cabot, J., Deruelle, L., Derras, M.: Xatkit: a multimodal low-code
  chatbot development framework. IEEE Access  \textbf{8},  15332--15346 (2020)

\bibitem{davis1973group}
Davis, J.H.: Group decision and social interaction: A theory of social decision
  schemes.  (1973)

\bibitem{deloach2007effectively}
DeLoach, S.B., Greenlaw, S.A.: Effectively moderating electronic discussions.
  The Journal of Economic Education  \textbf{38}(4),  419--434 (2007)

\bibitem{goda2014conversation}
Goda, Y., Yamada, M., Matsukawa, H., Hata, K., Yasunami, S.: Conversation with
  a chatbot before an online efl group discussion and the effects on critical
  thinking. The Journal of Information and Systems in Education
  \textbf{13}(1), ~1--7 (2014)

\bibitem{green1985rapid}
Green, P., Wei-Haas, L.: The rapid development of user interfaces: Experience
  with the wizard of oz method. In: Proceedings of the Human Factors Society
  Annual Meeting. vol.~29, pp. 470--474. SAGE Publications Sage CA: Los
  Angeles, CA (1985)

\bibitem{hamilton2018argument}
HAMILTON, L.A.C.F.L., Hunter, A., Potts, H.W.: Argument harvesting using
  chatbots. Computational Models of Argument: Proceedings of COMMA 2018
  \textbf{305}, ~149 (2018)

\bibitem{isbister2000helper}
Isbister, K., Nakanishi, H., Ishida, T., Nass, C.: Helper agent: Designing an
  assistant for human-human interaction in a virtual meeting space. In:
  Proceedings of the SIGCHI conference on Human Factors in Computing Systems.
  pp. 57--64 (2000)

\bibitem{janarthanam2017hands}
Janarthanam, S.: Hands-on chatbots and conversational UI development: build
  chatbots and voice user interfaces with Chatfuel, Dialogflow, Microsoft Bot
  Framework, Twilio, and Alexa Skills. Packt Publishing Ltd (2017)

\bibitem{kim2020bot}
Kim, S., Eun, J., Oh, C., Suh, B., Lee, J.: Bot in the bunch: Facilitating
  group chat discussion by improving efficiency and participation with a
  chatbot. In: Proceedings of the 2020 CHI Conference on Human Factors in
  Computing Systems. pp. 1--13 (2020)

\bibitem{lamm1973group}
Lamm, H., Trommsdorff, G.: Group versus individual performance on tasks
  requiring ideational proficiency (brainstorming): A review. European journal
  of social psychology  \textbf{3}(4),  361--388 (1973)

\bibitem{laughlin2011group}
Laughlin, P.R.: Group problem solving. Princeton University Press (2011)

\bibitem{lee2020solutionchat}
Lee, S.C., Song, J., Ko, E.Y., Park, S., Kim, J., Kim, J.: Solutionchat:
  Real-time moderator support for chat-based structured discussion. In:
  Proceedings of the 2020 CHI Conference on Human Factors in Computing Systems.
  pp. 1--12 (2020)

\bibitem{lorge1958survey}
Lorge, I., Fox, D., Davitz, J., Brenner, M.: A survey of studies contrasting
  the quality of group performance and individual performance, 1920-1957.
  Psychological bulletin  \textbf{55}(6), ~337 (1958)

\bibitem{mcleod1996ethnic}
McLeod, P.L., Lobel, S.A., Cox~Jr, T.H.: Ethnic diversity and creativity in
  small groups. Small group research  \textbf{27}(2),  248--264 (1996)

\bibitem{munsterberg1914mind}
M{\"u}nsterberg, H.: The mind of the juryman.  (1914)

\bibitem{nunamaker1991electronic}
Nunamaker, J., Dennis, A., Valacich, J., Vogel, D., George, J.: Electronic
  meeting systems to support group work. Commun. ACM  \textbf{34},  40--61 (01
  1991)

\bibitem{peng2019gremobot}
Peng, Z., Kim, T., Ma, X.: Gremobot: Exploring emotion regulation in group
  chat. In: Conference Companion Publication of the 2019 on Computer Supported
  Cooperative Work and Social Computing. pp. 335--340 (2019)

\bibitem{pinsonneault1989impact}
Pinsonneault, A., Kraemer, K.L.: The impact of technological support on groups:
  An assessment of the empirical research. Decision Support Systems
  \textbf{5}(2),  197--216 (1989)

\bibitem{shamekhi2018face}
Shamekhi, A., Liao, Q.V., Wang, D., Bellamy, R.K., Erickson, T.: Face value?
  exploring the effects of embodiment for a group facilitation agent. In:
  Proceedings of the 2018 CHI Conference on Human Factors in Computing Systems.
  pp. 1--13 (2018)

\bibitem{shaw1932comparison}
Shaw, M.E.: A comparison of individuals and small groups in the rational
  solution of complex problems. The American Journal of Psychology
  \textbf{44}(3),  491--504 (1932)

\bibitem{siemon2019one}
Siemon, D., Becker, F., Eckardt, L., Robra-Bissantz, S.: One for all and all
  for one-towards a framework for collaboration support systems. Education and
  Information Technologies  \textbf{24}(2),  1837--1861 (2019)

\bibitem{thorndike1938effect}
Thorndike, R.L.: The effect of discussion upon the correctness of group
  decisions, when the factor of majority influence is allowed for. The Journal
  of Social Psychology  \textbf{9}(3),  343--362 (1938)

\bibitem{van1997supporting}
Van~Genuchten, M., Cornelissen, W., Van~Dijk, C.: Supporting inspections with
  an electronic meeting system. Journal of Management Information Systems
  \textbf{14}(3),  165--178 (1997)

\bibitem{viller1991group}
Viller, S.: The group facilitator: a cscw perspective. In: Proceedings of the
  Second European Conference on Computer-Supported Cooperative Work ECSCW’91.
  pp. 81--95. Springer (1991)

\bibitem{zhao2014towards}
Zhao, R., Papangelis, A., Cassell, J.: Towards a dyadic computational model of
  rapport management for human-virtual agent interaction. In: International
  Conference on Intelligent Virtual Agents. pp. 514--527. Springer (2014)

\bibitem{zhou2020design}
Zhou, L., Gao, J., Li, D., Shum, H.Y.: The design and implementation of
  xiaoice, an empathetic social chatbot. Computational Linguistics
  \textbf{46}(1),  53--93 (2020)

\bibitem{ziller1962group}
Ziller, R.C., Behringer, R.D., Goodchilds, J.D.: Group creativity under
  conditions of success or failure and variations in group stability. Journal
  of Applied Psychology  \textbf{46}(1), ~43 (1962)

\end{thebibliography}
\bibliographystyle{splncs04}

\end{document}